\documentclass{ws-p9-75x6-50}
\usepackage{graphicx}
\usepackage{color}
\usepackage{amsbsy,amssymb,amsmath,bm,xspace}
\usepackage{fancybox}
\usepackage{fancyhdr}
\usepackage{rotating}
\def\one{{\mathchoice {\rm 1\mskip-4mu l} {\rm 1\mskip-4mu l} {\rm
1\mskip-4.5mu l} {\rm 1\mskip-5mu l}}}
 
\newcommand{\Integers}{\ensuremath{\mathbb{Z}}\xspace}

\begin{document}

\title{Hidden unity in the quantum description of matter}

\author{G. Ortiz and C.D. Batista}

\address{Theoretical Division, Los Alamos National Laboratory, 
USA \\E-mail: ortiz@viking.lanl.gov}

\maketitle

\abstracts{
We introduce an algebraic framework for interacting quantum systems
that enables studying complex phenomena, characterized by the
coexistence and competition of various broken symmetry states of
matter. The approach unveils the hidden unity behind seemingly
unrelated physical phenomena, thus establishing exact connections
between them. This leads to the fundamental concept of {\it
universality} of physical phenomena, a general concept not restricted
to the domain of critical behavior. Key to our framework is the concept
of {\it languages} and the construction of {\it dictionaries} relating
them.}

\section{The Dictionaries of Nature}

As Science progresses it tends to successfully describe the diverse
phenomena encountered in nature with fewer underlying principles.
Indeed, the search for the unifying principles behind the fundamental
laws of physics is a common theme in the life of a physicist and has a
very simple reason which is simplifying the understanding of the
universe in which we live. Even if we knew the ultimate laws that
govern the universe, could one predict the complex behavior observed in
nature?  This has been the subject of numerous works by very eminent
people, like Philip W. Anderson, who rightfully argued that the whole is
not necessarily the sum of its parts and  thus ``more is different."
\cite{anderson} It seems as if matter organizes in well-defined but
hard to decipher patterns.

The notion of symmetry has shaped our current conception of nature;
however, nature is also full of symmetry breakings. Therefore
understanding the idea of invariance and its corresponding conservation
laws is as fundamental as determining the causes that prevent such
harmony, and leads to more complex behavior. Unveiling and mastering the
organizing principles is important since it leads, for example, to the
design of new materials and devices with specific functionalities and
unprecedented technological applications. Who would not like to have a
room-temperature superconductor? However, the plethora of complex
phenomena exceeds our ability to explain them, in part because of a
lack of appropriate mathematical tools to disentangle its mysteries.
Since quantum complexity is characterized by the coexistence and
competition of various states of matter one needs an efficient and
well-controlled approach to these problems that goes beyond traditional
mean-field and semi-classical approximations.

Describing the structure and behavior of matter entails studying
systems of interacting quantum constituents (bosons, fermions, spins)
and essential to complexity are correlations, involving  non-linear
couplings, between their different components. In the
quantum-mechanical description of matter, each physical system is
naturally associated with a {\it language} of operators (for example,
quantum spin-1/2 operators) and thus to an algebra realizing this
language (e.g., the Pauli spin algebra generated by a family of
commuting quantum spin-1/2 operators). It is our point of view that
crucial to the successful understanding of the mechanisms driving
complexity is the realization of {\it dictionaries} (isomorphisms) 
connecting the different languages of nature and therefore
linking seemingly unrelated physical phenomena. The existence of
dictionaries provides not only a common ground to explore complexity
but leads naturally to the  fundamental concept of {\it universality},
meaning that different physical systems show the same behavior. In this
way, there is a concept of physical equivalence hidden in these
dictionaries.

In this chapter we present an algebraic framework for interacting
extended quantum systems that allows studying complex phenomena
characterized by the coexistence and competition of various broken
symmetry states of matter. We show that exact algebraic and group
theory methods are one of the most elegant and promising approaches
towards a complete understanding of quantum phases of matter and their
corresponding phase transitions. Previous to our work \cite{our1}
there were two seemingly unrelated examples of dictionaries: The
Jordan-Wigner (1928) \cite{jordan} and Matsubara-Matsuda
transformations (1956) \cite{matsu}. In addition to the generalization
of these  ($su(2)$) transformations to any irreducible spin
representation, spatial dimension and particle statistics, we have
proved a fundamental theorem which connects operators generating
different algebras (e.g., $su(D)$ spin-particle connections), unifying
the different languages known so far in the quantum-mechanical
description of matter. The chapter has been written with the intention
of providing the reader with the most fundamental concepts involved in
our algebraic framework and how they apply to study complex phenomena.
Much more details and examples can be found in the original manuscripts
\cite{our1,our3,our4}.

\section{Algebraic Approach to Interacting Quantum Systems}

The theory of operator algebras on Hilbert spaces was initiated by
Murray and von Neumann \cite{murray} as a tool to study unitary
representations of groups, and as a framework for a reformulation of
quantum mechanics. This area of research continued its development
independently in the realm of mathematical physics, and therefore
knowledge of those investigations remained bounded to 
specialists.  For use of $C^*$ and $W^*$ algebras as a framework for
quantum  statistical mechanics one can look at the books of Bratteli
and  Robinson \cite{bratteli}. For the purposes of our presentation one
only needs to have an elementary background in basic algebra
\cite{algebra}, and specially group theory \cite{cornwell}. In
particular, Lie algebras and groups.

Here we are concerned with quantum lattice systems. A quantum lattice
is identified with $\Integers^{N_s}$, where $N_s$ is the total number
of lattice sites (or modes). Associated to each lattice site ${\bf j}
\in \Integers^{N_s}$ there is a Hilbert space ${\cal H}_{\bf j}$ of
finite dimension $D$ describing the ``local'' modes. The total Hilbert
space is ${\cal H} = \bigotimes_{\bf j} {\cal H}_{\bf j}$.  Since we
are mostly interested in zero temperature properties, a state of the
system is simply a vector $| \Psi \rangle$ in ${\cal H}$, and an
observable is a selfadjoint operator $O: {\cal H} \rightarrow {\cal
H}$. The dynamical evolution of the system is determined by its
Hamiltonian $H$. The topology of the lattice, dictated by the
connectivity and range of the interactions in $H$, is an important
element in establishing complexity. In the case of quantum continuous
systems we can still use the present formalism after discretizing the
space. Going beyond this approach is outside the scope of these notes.

\begin{figure}[htb]
\hspace*{1.2cm}
\includegraphics[angle=0,width=10.0cm,scale=1.0]{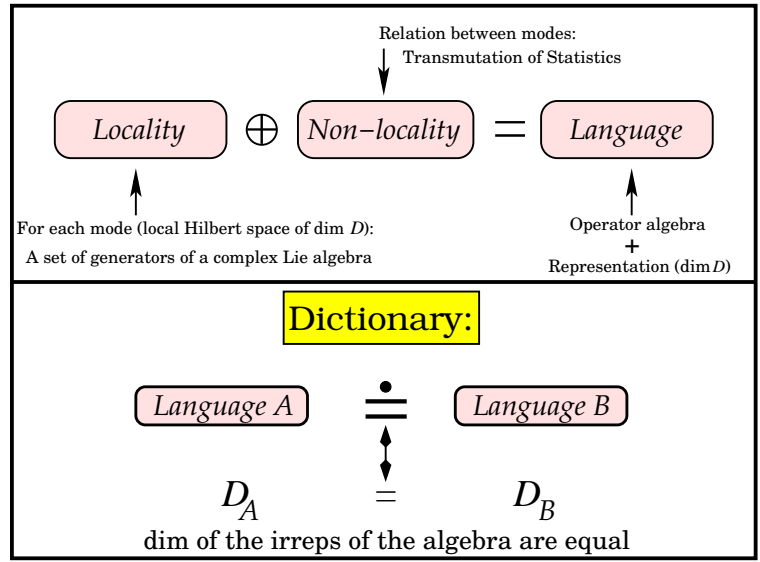}
\caption{
Definition of a {\it language} and theorem behind the construction of
the dictionaries of nature. In the upper panel we show schematically
what elements define a {\it language }$={\cal A}\wedge \Gamma_A$, where
${\cal A}$ is the algebra and $\Gamma_A$ a particular irrep. In the
lower panel we establish the criteria to build a dictionary, given two
languages $A$ and $B$. This criteria is based upon Burnside's
fundamental theorem of algebra \cite{algebra}.}
\label{fig1}
\end{figure}

As mentioned above, each physical system is naturally associated with a
language of operators, and thus to an algebra realizing this language.
Formally, a {\it language} is defined by an operator algebra and a
specific representation of the algebra. Mathematically, we use the
following notation: {\it language} = ${\cal A}\wedge \Gamma_A$, where
${\cal A}$ is the operator algebra and $\Gamma_A$ is a particular
irreducible representation (irrep) of the local algebra ${\cal A}_{\bf
j}$ associated to ${\cal A}$, of dimension dim $\Gamma_A = D$ (see  Fig.
\ref{fig1}). 

For the sake of clarity, let us choose the phenomenon of magnetism to
illustrate the key ideas. This is one of the most intriguing and not
fully understood problems in condensed matter physics where strong
correlations between electrons (of electrostatic origin) are believed
to be the essence of the problem. To describe the phenomenon using a
minimal model (i.e., a model that only includes the relevant degrees of
freedom) distinct approaches can be advocated depending upon the
itineracy of the electrons that participate in the magnetic processes.
In one extreme (e.g., insulators) a description in terms of localized
quantum spins is appropriate, while in the other (e.g., metals)
delocalization of the electrons is decisive and cannot be ignored. We
immediately identify the languages associated to each description:
quantum spins (e.g., Pauli algebra) and fermions (spin-1/2 Fermi
algebra). Are these really different descriptions? Is
there a dictionary that may connect the two languages? Let's assume
that we decide to use the quantum spins language. What other seemingly
unrelated phenomena are connected to magnetism?  Can we relate phases
of matter corresponding to dissimilar phenomena?  Can an arbitrary
physical system be mapped, for instance, onto a pure magnetic system
(an array of quantum spins)? 

In the following we will answer these questions by examples. A
fundamental concept of universality, complementary to the one used in
critical phenomena, emerges as a consequence of unveiling the hidden
unity in the quantum-mechanical description of matter. 

\subsection{Building the Dictionaries of Nature}

\begin{figure}[htb]
\hspace*{2.7cm} \includegraphics[angle=0,width=7.cm,scale=1.0]{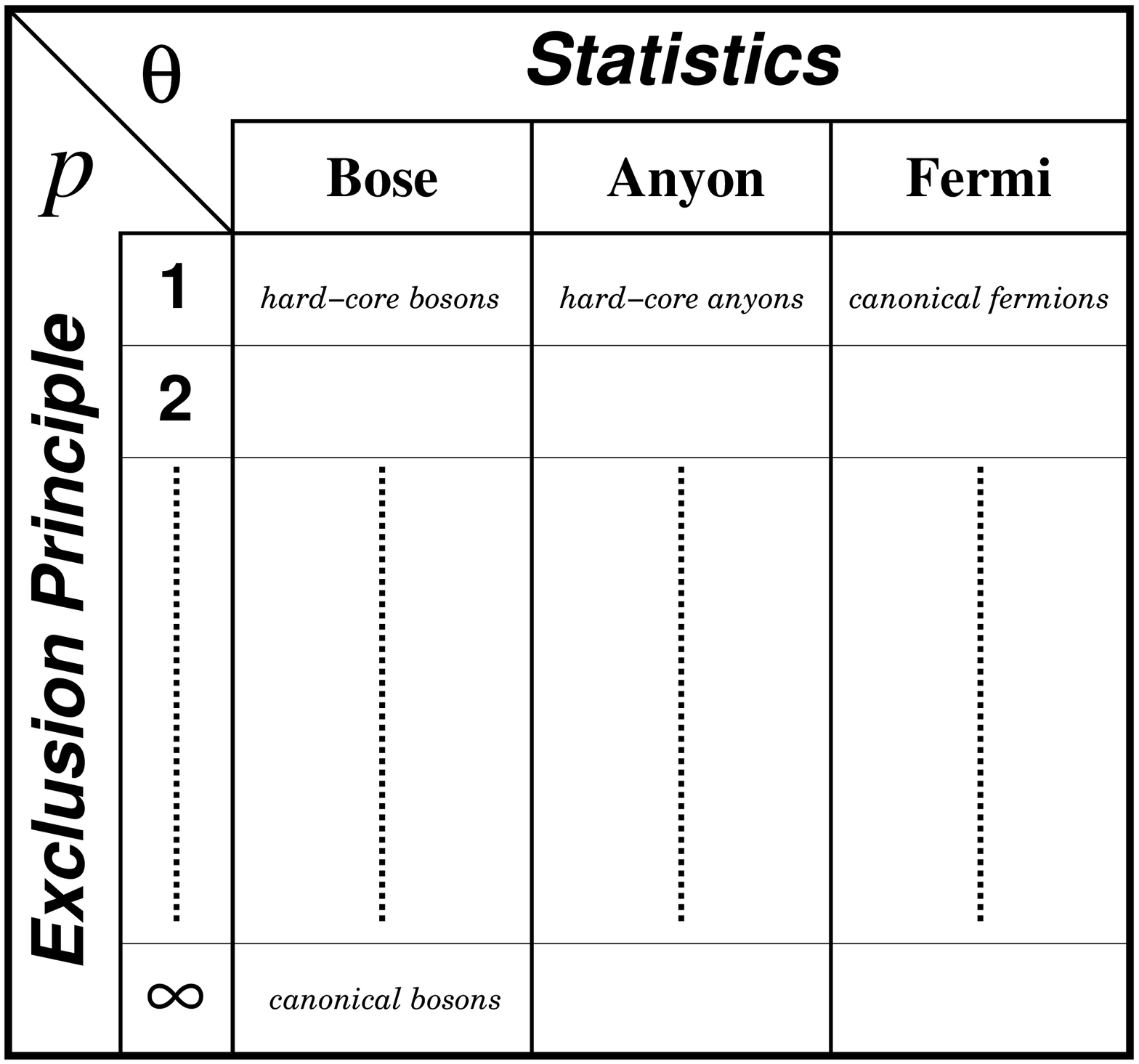}
\caption{
Classification of single-flavor ``particles'' according to the 
fundamental exclusion ($p$) and exchange statistics ($\theta$)
principles. Here we only consider double-operator algebras;  the
general case of para-statistics (triple-operator algebras) is excluded.
Each raw represents an equivalence class and we have used as
representatives of those equivalence classes the known particles found 
in nature. Most fundamental is the concept of {\it language} which
uniquely defines the type of particle.}
\label{fig2}
\end{figure}

The simplest model of magnetism is provided by the Heisenberg-Ising
Hamiltonian
\begin{equation}
H = J\sum_{j=1}^{N_s-1} \Delta \ S_j^z S_{j+1}^z + 
\frac{1}{2}\ (S_j^+ S_{j+1}^- + S_j^- S_{j+1}^+) \ ,
\label{heisenberg}
\end{equation}
where the operators $S^\mu_j$ satisfy an $\bigoplus_j su(2)$ algebra
($SU(2)$ symmetry of the Hamiltonian is recovered at the points
$\Delta=\pm1$). If we work in the $S=1/2$ irrep it is  well-known that
the model can be mapped onto an interacting spinless fermion model
through the Jordan-Wigner transformation \cite{jordan}. More generally,
the model can mapped onto \cite{our1,our4}
\begin{equation}
H=J\sum_{j=1}^{N_s-1} \Delta (n_j - \frac{1}{2})(n_{j+1} -
\frac{1}{2}) + \frac{1}{2} \ (a^{\dagger}_j a^{\;}_{j+1} +
a^{\dagger}_{j+1}a^{\;}_{j})
\end{equation}
through the transformation ($n_j=a^{\dagger}_j a^{\;}_{j}$ and $0 \leq
\theta < 2\pi$)
\begin{eqnarray}
\begin{cases}
S^+_{j}={a}^\dagger_{j} \, K_{j}(\theta) \\
S^-_{j}= K_{j}^\dagger(\theta) \, {a}^{\;}_{j}\\
S^z_{j}= n_{j} - \frac{1}{2} 
\end{cases} \ ,
\end{eqnarray}
where the non-local statistical operator or transmutator
$K_{j}(\theta)$ is given by
\begin{equation}
K_{j}(\theta)=e^{i \theta  \sum_{{i}<{j}} n_{i}} = 
\prod_{{i}<{j}} [1+(e^{i\theta}-1) \ n_{i}] 
\end{equation}
since $n_{j}^2=n_{j}$ (for any $\theta$), and satisfy $ K_{j}(\theta)
K^\dagger_{j}(\theta)=K^\dagger_{j}(\theta)  K_{j}(\theta)=\one$. In
this way we transformed the original localized spin-1/2 problem into an
itinerant gas of (anyon) particles obeying the double-operator algebra
($[A,B]_\theta=AB - e^{i\theta} BA$)
\begin{eqnarray}
\begin{cases}
[ a^{\;}_i,a^{\;}_j ]_\theta = 
[ a^\dagger_i,a^\dagger_j ]_\theta=0  
 \ , \\
{[}a^{\;}_i,a^\dagger_j{]}_{-\theta}=\delta_{ij} (1-(e^{-i
\theta}+1)n_j)
\; , \;
[ n_i, a^\dagger_j ]= \delta_{ij} 
a^\dagger_j  \ ,
\end{cases}
\end{eqnarray}
(for $i \leq j$). Each statistical angle $\theta$ provides a different
particle language and defines the exchange statistics of the particles.
The case $\theta=\pi$ corresponds to canonical spinless fermions
\cite{jordan} while $\theta=0$ represents hard-core (HC) bosons
\cite{matsu}. In all cases one can accommodate up to a single particle
($p=1$) per quantum state, $(a^\dagger_j)^{p+1}=0$ (i.e, the particles
are ``HC''). Figure \ref{fig2} shows a classification of particles
according to the independent concepts of exchange statistics and
generalized Pauli exclusion principle \cite{our4}. The statistical
operator connects particles within each equivalence class. 

In order to construct a dictionary one also needs the inverse mapping
\begin{eqnarray}
\begin{cases}
\displaystyle
{a}^\dagger_{j}=\prod_{{i}<{j}} 
[\frac{e^{-i\theta}+1}{2}+(e^{-i\theta}-1) \ S^z_i]\
S^+_{j}\\
\displaystyle
{a}^{\;}_{j}=\prod_{{i}<{j}} 
[\frac{e^{i\theta}+1}{2}+(e^{i\theta}-1)\ S^z_i] \
S^-_{j} \\ n_{j}=S^z_j +\frac{1}{2} 
\end{cases} \ .
\end{eqnarray}

Let us consider now the same Hamiltonian, Eq. (\ref{heisenberg}), but
with spin operators in an $S=1$ irrep. A possible mapping in terms of
two-flavor (or $s$=1/2) particles is \cite{our1}
\begin{eqnarray}
\begin{cases}
S^+_{j}=\sqrt{2} \ ({a}^\dagger_{j\uparrow} \, K_{j}(\theta) +
K_{j}^\dagger(\theta) \, {a}^{\;}_{j\downarrow}) \\
S^-_{j}=\sqrt{2} \ (K_{j}^\dagger(\theta) \, {a}^{\;}_{j\uparrow} + 
{a}^\dagger_{j\downarrow} \, K_{j}(\theta))\\
S^z_{j}= n_{j\uparrow} - n_{j\downarrow} 
\label{HC}
\end{cases} \ ,
\end{eqnarray}
where the non-local transmutator ($n_{j\alpha}=a^\dagger_{j\alpha}
a^{\;}_{j\alpha}$, $n_{j}={n}_{j\uparrow}+{n}_{j\downarrow}$, and 
$\alpha=\uparrow,\downarrow$)
\begin{equation}
K_{j}(\theta)=e^{i \theta  \sum_{{i}<{j}} n_{i}} = 
\prod_{{i}<{j}} [1+(e^{i\theta}-1) \ n_{i}] 
\end{equation}
(${n}_{j\alpha}{n}_{j\beta}=\delta_{\alpha \beta} \, {n}_{j\alpha}$)
allows rotation of the statistics of the particles whose algebra is 
determined by ($i \leq j$)
\begin{eqnarray}
\begin{cases}
{[} a^{\;}_{i\alpha}, a^{\;}_{j\beta} {]}_\theta =
{[} a^\dagger_{i\alpha}, a^\dagger_{j\beta} {]}_\theta = 0  \ , \\
{[} a^{\;}_{i\alpha}, a^\dagger_{j\beta} {]}_{-\theta}
= \delta_{ij}\begin{cases}
\mbox{$\displaystyle 1-e^{-i\theta}\, {n}_{j\alpha} -
{n}_{j}$}&
		\text{if $\alpha = \beta$}, \\
                -e^{-i\theta} \, a^\dagger_{j\beta} a^{\;}_{j\alpha} & 
		\text{if $\alpha \neq \beta$} , 
		\end{cases}   
\end{cases} \ .
\end{eqnarray}
In this case a more restrictive version of the Pauli exclusion
principle applies where one can accommodate no more than a single
particle per site regardless of $\alpha$, i.e., 
$a^\dagger_{j\alpha}a^\dagger_{j\beta}=0 , \ \forall (\alpha, \beta)$.
In this language the resulting Hamiltonian is \cite{Note1}
\begin{equation}
H=J \sum_{j=1}^{N_s-1} \!\!  \Delta (n_{j\uparrow} -
n_{j\downarrow}) (n_{j+1 \uparrow} - n_{j+1 \downarrow}) + 
\sum_\alpha (a^{\dagger}_{j\alpha} a^{\;}_{j+1 \alpha} +
a^{\dagger}_{j\alpha}a^{\dagger}_{j+1 \bar{\alpha}} + {\rm H.c.}) \ .
\end{equation}
One can define an inverse mapping as in the $S=1/2$ case, thus building
the appropriate dictionary \cite{our1}. In a similar fashion, one could
continue for higher spin $S$ irreps and would find that HC particles
have $N_f=2S$ flavors (we call these generalized Jordan-Wigner
particles) \cite{our1}. Of course, this is not the only way to proceed.
For example, for half-odd integer cases where $2S+1 = 2^{\bar{N}_f}$ a
simple transformation in terms of standard canonical multiflavor
fermions is possible \cite{our1,our4}.

What is the unifying concept behind the construction of the
dictionaries of nature? When is it possible to build a dictionary
between two arbitrary languages? The answers lie in the application of
the following theorem together with the transmutation of statistics
\cite{our4} (see lower panel in Fig. \ref{fig1}):

\noindent
{\it Fundamental Theorem:} Given a set of generators of a complex Lie
algebra $\tilde{\cal L}$ in a particular ($D$-dimensional) $\Gamma_A$ 
irrep, it is always possible to write these generators as a function of
the identity and the generators of a bosonic language ${\cal B}\wedge
\Gamma_B$  where dim $\Gamma_B =D$. Similarly, each generator of the
bosonic language can be written in terms of the identity and the
generators of $\tilde{\cal L}$ in the $\Gamma_A$ irrep. The algebra
defining a bosonic language is ${\cal B} = \bigoplus_{\bf j} {\cal
B}_{\bf j}$, where the local algebra ${\cal B}_{\bf j}$ is a Lie algebra.
 
This theorem provides the necessary and sufficient conditions to
connect two bosonic languages. To construct dictionaries (isomorphisms)
for arbitrary (bosonic, fermionic or anyonic) languages one needs to
complement the theorem with the transmutators of statistics
\cite{our4}. In our first example ${\rm dim} \ {\cal H}_j=D=2$ and 
the bosonic language is $\bigoplus_j su(2) \wedge S=\frac{1}{2}$, the
fermionic and anyonic languages are obtained after application of the 
statistical transmutator. The second example only differs in the
representation ($S=1, D=3$). 

In this section the emphasis was put in establishing the commonalities
between the different languages. In addition, to illustrate the notion
of coexistence and competition of phases one needs the solution of the
model. In the next section we will explain the concept of universality 
by realizing a dictionary that connects different Lie algebras with
$D=3$ representations.

\subsection{Unveiling Order behind Complexity}

The field of quantum phase transitions studies the changes that can
occur in the macroscopic properties of matter at zero temperature due
to changes in the parameters characterizing the system. To identify
broken symmetry phases (ground states (GSs)) what is typically used as
a working principle is Landau's postulate of an order parameter (OP).
One of the major properties of this OP is the symmetry rules it obeys.
The space in which it resides is often represented as the quotient set
of the symmetry group of the disordered phase and the symmetry group of
the ordered phase. While one generally knows what to do if the OP is
known, Landau's postulate gives no procedure for finding it. In this
section we describe a simple algebraic framework for identifying OPs.

The system we want to study is the $SU(2)$-invariant model Hamiltonian 
($J<0$)
\begin{equation} 
H_\vartheta = J\sqrt{2} \sum_{\langle {\bf i},{\bf j} \rangle} \left [
\cos \vartheta \ {\bf S}_{\bf i} \cdot {\bf S}_{{\bf j}} + \sin \vartheta
\left ( {\bf S}_{\bf i} \cdot {\bf S}_{{\bf j}} \right )^2 \right ] ,
\label{spin1}
\end{equation}
where ${\bf S}_{\bf j}$ is a $S$=1 operator satisfying the algebra 
$\bigoplus_{\bf j} su(2)$, as before. Summation is over bonds $\langle
{\bf i},{\bf j} \rangle$ of a regular $d$-dimensional lattice with
$N_s$ sites and coordination ${\sf z}$. As we will see, for certain
values of $\vartheta$ this  Hamiltonian is highly symmetric: for
$\vartheta = \frac{\pi}{4}$ and  $\frac{5\pi}{4}$ it is explicitly
invariant under uniform $SU(3)$ transformations on the spins, while for
$\vartheta=\pm \frac{\pi}{2}$, it is explicitly invariant under
staggered conjugate rotations of the two sublattices \cite{our3}. 

The case $\vartheta=\frac{\pi}{4}$ can be conveniently written in the
($s=1/2$) HC boson language of Eq. (\ref{HC}) with $\theta=0$
\begin{equation}
H_{\frac{\pi}{4}}=J  \sum_{\langle {\bf i},{\bf j} \rangle}\!
\sum_\alpha \left ( a^\dagger_{{\bf i} \alpha} a^{\;}_{{\bf j} \alpha}
+ {\rm H.c.} \right ) + 2   {\bf s}_{\bf i} \cdot  {\bf s}_{\bf j} + 2 
\left (1 - \frac{{{n}}_{\bf i}+{{n}}_{\bf j}}{2} + \frac{3}{4}
{{n}}_{\bf i} {{n}}_{\bf j}  \right ) \ ,
\label{hamilt}
\end{equation}
where ${\bf s}_{\bf j}=\frac{1}{2} a^\dagger_{{\bf j} \mu}
\boldsymbol{\sigma}_{\mu \nu} a^{\;}_{{\bf j} \nu}$ with
$\boldsymbol{\sigma}$ denoting Pauli matrices.

For a system of ${\cal N}={\cal N}_{\uparrow}+{\cal N}_{\downarrow}$  
(${\cal N}\leq N_s$) HC bosons the exact GS is \cite{our3}
\begin{equation}\label{GS}
| \Psi_0({\cal N},S_z) \rangle = (\tilde{a}^{\dagger}_{{\bf 0}
\uparrow})^{{\cal N}_{\uparrow}} (\tilde{a}^{\dagger}_{{\bf
0}\downarrow})^{{\cal N}_{\downarrow}} | 0 \rangle \ ,
\end{equation}
with an energy $E_0/N_s=J {\sf z}$ and a total $S_z = \frac{{\cal
N}_{\uparrow}-{\cal N}_{\downarrow}}{2}$. The operator
$\tilde{a}^{\dagger}_{{\bf 0}\alpha}$ is the ${\bf k}={\bf 0}$
component of  ${a}^\dagger_{{\bf j}\alpha}$, i.e.,
$\tilde{a}^\dagger_{{\bf k} \alpha}= \frac{1}{\sqrt{N_s}} \sum_{\bf j}
\exp[i{\bf k}\cdot {\bf r}_{\bf j}] \ {{a}}^\dagger_{{\bf j}
\alpha}$. The quasihole and quasiparticle excited states are
\begin{eqnarray}
                \begin{cases}
| \Psi^h_{\bf k}({\cal N},S_z) \rangle = \tilde{a}^{\;}_{{\bf k}
\alpha} | \Psi_0({\cal N},S_z)\rangle &
        \text{quasihole}, \\
| \Psi^p_{\bf k}({\cal N},S_z) \rangle = \tilde{a}^{\dagger}_{{\bf
k} \alpha} | \Psi_0({\cal N},S_z)\rangle &
        \text{quasiparticle} ,
        \end{cases}
\end{eqnarray}
with the excitation energy of each being $\omega_{\bf k} = J{\sf z} 
(\frac{1}{\sf z} \sum_{\nu} e^{i {\bf k} \cdot {\bf e}_\nu}-1)$ where the
sum runs over the vectors ${\bf e}_\nu$ which connect a given site to
its ${\sf z}$ nearest neighbors. In the  $|{\bf k}| \rightarrow 0$
limit, $\omega_{\bf k}\rightarrow 0$.

Clearly the GS in Eq.~(\ref{GS}) is a ferromagnetic Bose-Einstein (BE)
condensate with arbitrary spin polarization, and the form of the result
is {\it independent of the spatial dimensionality of the lattice}. We
note that different values of $S_z$ correspond to the different
orientations of the magnetization ${\cal M}$ associated to the broken
$SU(2)$ spin rotational symmetry of the GS. We also note that the
degeneracy of states with different number of particles ${\cal N}$
indicates a broken $U(1)$ charge symmetry (conservation of the number
of particles) associated to the BE condensate. A signature of Bose
condensation is the existence of off-diagonal long-range order (ODLRO) 
in the correlation function $\Phi_{\alpha \beta}({\bf ij})=\langle 
{{a}}^\dagger_{{\bf i} \alpha} {{a}}^{\;}_{{\bf j}\beta} \rangle$ since
that implies that there is at least one eigenvector 
with an eigenvalue of order $N_s$ \cite{Note2}.

We can easily compute the magnetization ${\cal M}$ and phase coherence
of these various (non-normalized) degenerate GSs for a given density
$\rho=\frac{{\cal N}}{N_s}$.  For example, in the fully polarized case,
${\cal N}={\cal N}_\uparrow$, ${\cal M}=\langle S^z_{\bf j} \rangle =
\rho$, and the ODLRO (${\bf r}_{\bf i} \neq {\bf r}_{\bf j}$)
$\Phi_{\uparrow \uparrow}({\bf ij})= \frac{\rho (1-\rho)}{1-\epsilon}$,
where $\epsilon=1/N_s$.  Similarly, the two-particle correlation
function $\langle \Delta^\dagger_{\bf i} \Delta^{\;}_{\bf j} \rangle =
\Phi_{\uparrow \uparrow}({\bf ij})
\frac{(\rho-\epsilon)(1-\rho-\epsilon)}{(1-2\epsilon) (1-3\epsilon)}$,
where $\Delta^\dagger_{\bf i}={{a}}^\dagger_{{\bf i}
\uparrow}{{a}}^\dagger_{{\bf i}+\boldsymbol{\delta} \uparrow}$
\cite{our3}. 

The exact solution defines the features of the phase diagram that our
proposed framework must qualitatively admit. We will see now that both
OPs (magnetization and phase) are embedded in an $SU(3)$ OP. To this
end one introduces a new language based upon the $\bigoplus_{\bf j}
su(3)$ algebra in the fundamental representation with generators
satisfying  ($\mu,\nu \in [0,2]$) $[{\cal S}^{\mu \mu'}({\bf j}),{\cal
S}^{\nu \nu'}({\bf j})]= \delta_{\mu' \nu} {\cal S}^{\mu \nu'}({\bf
j})-\delta_{\mu \nu'} {\cal S}^{\nu \mu'}({\bf j})$. One can rewrite 
Eq. (\ref{spin1}) (up to an irrelevant constant) in this new language as
\begin{eqnarray}
H_\vartheta= J \sqrt{2}\sum_{\langle {\bf i},{\bf j} \rangle} \left
[\cos \vartheta \ {\cal S}^{\mu \nu}({\bf i}) {\cal S}^{\nu \mu}({\bf
j}) + (\sin \vartheta - \cos \vartheta) \ {\cal S}^{\mu \nu}({\bf i})
\tilde{\cal S}^{\nu \mu}({\bf j}) \right ] \ ,
\end{eqnarray}
where ${\cal S}({\bf j})$ defines the $su(3)$ spin-particle mapping 
\begin{equation}
{\cal S}({\bf j})= \begin{pmatrix} \frac{2}{3} - {n}_{{\bf j}}
&a^{\;}_{{\bf j} \uparrow} & a^{\;}_{{\bf j} \downarrow} \\
a^\dagger_{{\bf j} \uparrow}&{n}_{{\bf j} \uparrow} -\frac{1}{3}&
a^\dagger_{{\bf j} \uparrow} a^{\;}_{{\bf j} \downarrow} \\
a^\dagger_{{\bf j} \downarrow}&a^\dagger_{{\bf j} \downarrow} a^{\;}_{{\bf
j} \uparrow}&{n}_{{\bf j} \downarrow}-\frac{1}{3} \end{pmatrix} \ , \ \
\tilde{\cal S}({\bf j})= \begin{pmatrix} \frac{2}{3} - {n}_{{\bf j}}
&-a^\dagger_{{\bf j} \downarrow} & -a^\dagger_{{\bf j} \uparrow} \\
-a^{\;}_{{\bf j} \downarrow}&{n}_{{\bf j} \downarrow}
-\frac{1}{3}& a^\dagger_{{\bf j} \uparrow} a^{\;}_{{\bf j} \downarrow}
\\ -a^{\;}_{{\bf j} \uparrow}&a^\dagger_{{\bf j} \downarrow}
a^{\;}_{{\bf j} \uparrow}&{n}_{{\bf j} \uparrow}-\frac{1}{3} 
\end{pmatrix}\ .
\label{spinsu3}
\end{equation}
$\tilde{\cal S}({\bf j})$ generates the conjugate representation.
For  $\vartheta=\frac{\pi}{4}$ we easily recognize the $SU(3)$
symmetric Heisenberg model and when $J<0$ the GS is the state with
maximum total $SU(3)$ spin ${\cal S}$. The OP associated with this broken
symmetry is the total $SU(3)$ magnetization 
\begin{equation}
{\cal S}^{\mu \nu }({\bf k})=\frac{1}{N_s}
\sum_{\bf j} e^{i {\bf k} \cdot {\bf r_j}} \ {\cal S}^{\mu \nu }({\bf j})
\end{equation}
which has eight independent components. When $\langle {\cal S}^{\mu \nu
} \rangle \neq 0$, the system orders, and the coexistence of a
ferromagnetic phase and a BE condensation becomes more evident: In the
HC boson language both OPs correspond to different components of the
$SU(3)$ OP (see Eq. (\ref{spinsu3})). Table \ref{tab:op} summarizes the
relations between OPs and quantum phases in the different
languages for the homogeneous ${\bf k}={\bf 0}$ case.

\begin{table}[htb]
\caption{Generators of OPs and its relations for three
different languages ${\cal A}\wedge \Gamma_A$. Each column represents a
language, in this case dim $\Gamma_A= D =3$. M stands for
magnetism, SN spin-nematic, BE Bose-Einstein condensation, and CDW
charge-density wave. $su(3) \wedge$ FR is the hierarchical language
with FR meaning fundamental representation.}
\label{tab:op}
\begin{center}
\footnotesize
\begin{tabular}{|c|c|c|}
\hline
\raisebox{0pt}[13pt][7pt]{\large $su(2) \wedge S=1$} &
\raisebox{0pt}[13pt][7pt]{\large $\mbox{HC bosons} \wedge \alpha =2$} &
\raisebox{0pt}[13pt][7pt]{\large $su(3) \wedge \mbox{FR}$}\\ 
\hline \hline
\begin{minipage}{2.3in}
\begin{eqnarray}
\hspace*{-0.2cm}
&{\rm M}&
\begin{cases}
\vspace*{0.2cm}
S^x=\frac{1}{\sqrt{2}} ({\cal S}^{0 1}+{\cal S}^{2 0}+ {\cal S}^{0
2}+{\cal S}^{1 0}) \\
\vspace*{0.2cm}
S^y=\frac{-1}{\sqrt{2}i} ({\cal S}^{0 1}+{\cal S}^{2 0}-{\cal S}^{0
2}-{\cal S}^{1 0})   \\
\vspace*{0.2cm}
S^z={\cal S}^{1 1}-{\cal S}^{2 2} 
\nonumber \\
\end{cases}
\\
\hspace*{-0.2cm}
&{\rm SN}&
\begin{cases}
\vspace*{0.2cm}
(S^x)^2=\frac{2}{3}+\frac{1}{2}({\cal S}^{1 2}+{\cal S}^{2 1}+{\cal
S}^{0 0})  \\
\vspace*{0.2cm}
(S^z)^2=\frac{2}{3}-{\cal S}^{0 0}   \\
\vspace*{0.2cm}
\left \{ S^x,S^y \right \}=i ({\cal S}^{2 1}-{\cal S}^{1 2}) \\
\vspace*{0.2cm}
\left \{ S^x,S^z \right \}=\frac{1}{\sqrt{2}} ({\cal S}^{0 1}-{\cal
S}^{2 0}-{\cal S}^{0 2}+{\cal S}^{1 0})   \\
\vspace*{0.2cm}
\left \{ S^y,S^z \right \}=\frac{-1}{\sqrt{2}i} ({\cal S}^{0 1}-{\cal
S}^{2 0}+{\cal S}^{0 2}-{\cal S}^{1 0}) \nonumber
\end{cases}
\end{eqnarray}
\end{minipage} 
 & 
\begin{minipage}{1.4in}
\begin{eqnarray}
\hspace*{-0.2cm}
&{\rm M}&
\begin{cases}
\vspace*{0.2cm}
s^x= \frac{1}{2}({\cal S}^{1 2}+{\cal S}^{2 1})  \\
\vspace*{0.2cm}
s^y= \frac{1}{2i}({\cal S}^{1 2}-{\cal S}^{2 1})  \\
\vspace*{0.2cm}
s^z= \frac{1}{2}({\cal S}^{1 1}-{\cal S}^{2 2}) \nonumber \\
\end{cases}
\\
\hspace*{-0.2cm}
&{\rm BE}&
\begin{cases}
\vspace*{0.2cm}
{a}^{\dagger}_{\uparrow}={\cal S}^{1 0}  \\
\vspace*{0.2cm}
{a}^{\dagger}_{\downarrow}={\cal S}^{2 0}  \\
\vspace*{0.2cm}
{a}^{\;}_{\uparrow}={\cal S}^{0 1}   \\
\vspace*{0.2cm}
{a}^{\;}_{\downarrow}={\cal S}^{0 2} \nonumber \\
\end{cases}
\\
\hspace*{-0.2cm}
&{\rm CDW}&
\begin{cases}
{n}=\frac{2}{3}- {\cal S}^{00} \nonumber 
\end{cases}
\end{eqnarray}
\end{minipage} 
& 
\begin{minipage}{0.6in}
\begin{center}
${\cal S}^{\mu\nu}$ \\ 
\vspace*{0.3cm}
$\mu,\nu \in [0,2]$ \\ 
\vspace*{0.3cm}
${\rm Tr } \, {\cal S} =0 $
\end{center}
\end{minipage} 
\\ 
\hline
\end{tabular}
\end{center}
\end{table}

A concept of universality naturally emerges from the dictionaries:
Many apparently different problems in nature have the same underlying
algebraic structure and, therefore, the same physical behavior. If it
is the whole system Hamiltonian that maps onto another in a different
language (like the example we described above), the universality
applies to all length and time scales. However, sometimes only
particular invariant subspaces of the original Hamiltonian map onto
another system Hamiltonian. In this case, universality is only
manifested at certain energy scales. The $t$-$J_z$ chain model provides
a beautiful example of the latter situation \cite{our2}: the low-energy
manifold of states maps onto an $XXZ$ model Hamiltonian, which can be
exactly solved using the Bethe ansatz.  

The fact that two dissimilar physical phenomena share the same set of
critical exponents is also known as universality. Those critical
phenomena are grouped into universality classes. Members of a given
universality class have the same broken symmetry group (OPs), and the
long-wavelenght excitations are described by a unique fixed-point
Hamiltonian. The idea of universality that emerges from our work is
complementary to the one used to analyze critical behavior. It is not
restricted to the study of critical phenomena, but can be exploited in
conjunction with Renormalization  Group techniques.

\section{Concluding Remarks}

We presented an algebraic framework aimed at uncovering the order
behind the potential multiplicity of complex phases in interacting
quantum systems, a new paradigm at the frontiers of condensed matter
physics. Crucial to this approach is the existence of dictionaries
(isomorphisms) that permits to connect the different languages used in
the quantum-mechanical description of matter. We also introduced the
idea of universality of physical phenomena, a concept that naturally
emerges from those dictionaries. In all cases we have given precise
mathematical definitions to these physical terms.

The development of exact algebraic methods is one of the most elegant
and promising tools towards the complete understanding of quantum
phases of matter and their corresponding phase transitions. Often these
systems are near quantum criticality which makes their study extremely
complicated, if not impossible, for the traditional techniques such as
mean-field or perturbative schemes. Precisely the same reason which
prevents the use of these theories is the key for the successful
application of algebraic methods: The absence of a small parameter and
degeneracy for different quantum complex orderings. 

There are several reasons why our algebraic framework constitutes a
powerful method to study complex phenomena in interacting quantum
systems. Most importantly: To connect seemingly unrelated physical
phenomena (e.g., models for high-$T_c$ or heavy fermion systems and 
quantum spin theories); identify the general symmetry principles behind
complex phase diagrams; unveil hidden symmetries (and associated order
parameters) to explore new states of matter with internal orders not
envisaged before; obtain exact solutions of relevant physical models
that display complex ordering at certain points in Hamiltonian space.

The algebraic framework for identifying OPs and possible broken
symmetry phases of quantum systems can be summarized as follows
\cite{our4}:

\noindent
$\Large \color{black} \bullet$
Identify the dimension $D$ of the local Hilbert space ${\cal H}_{\bf j}$ 
which sets the dimension of the irrep $\Gamma_A$ associated to the
language A.

\noindent
$\Large \color{black} \bullet$
The OP is constructed from a hierarchical group. A hierarchical
language is one whose local algebra ${\cal A}_{\bf j}$ has as
fundamental representation of dimension $D$. 

\noindent
$\Large \color{black} \bullet$
Identify the embedded subgroups. Reduce the fundamental representation
of the hierarchical group according to the irreps of the embedded
subgroups, thereby establishing a hierarchical classification of the
OPs and an enumeration of the possible broken symmetry phases.

%


\begin{thebibliography}{99}

\bibitem{anderson}
P.W. Anderson, \Journal{\Sci}{177}{393}{1972}.

\bibitem{jordan}
P. Jordan and E. Wigner, \Journal{\ZP}{47}{631}{1928}.

\bibitem{matsu}
T. Matsubara and H. Matsuda, \Journal{\PTP}{16}{569}{1956}.

\bibitem{our1}
C.D. Batista and G. Ortiz, \Journal{\PRL}{86}{1082}{2001}; {\em 
Condensed Matter Theories}, Vol. 16.

\bibitem{our3}
C.D. Batista, G. Ortiz, and J.E. Gubernatis, {\em Unveiling Order
behind Complexity: Coexistence of Ferromagnetism and Bose-Einstein
Condensation}, unpublished preprint.

\bibitem{our4}
C.D. Batista and G. Ortiz, {\em Algebraic Approach to Interacting Quantum
Systems}, unpublished preprint.

\bibitem{murray}
F.J. Murray and J. von Neumann, {\em Ann. Math.} {\bf 37}, 116 (1936).

\bibitem{bratteli}
O. Bratteli and D.W. Robinson, {\em Operator Algebras and Quantum
Statistical Mechanics}, Vols. I and II (Springer-Verlag, New York,
1987-97).

\bibitem{algebra}
N. Jacobson, {\em Basic Algebra}, Vols. I and II (W.H. Freeman, New York,
1985-89).

\bibitem{cornwell}
J.F. Cornwell, {\em Group Theory in Physics} (Academic Press, San Diego,
1997).

\bibitem{Note1}
At this point two comments are in order. The first refers to the
irrelevance of particle statistics for one-dimensional models with
nearest-neighbor interactions like Eq. (\ref{heisenberg}). Clearly one
needs two or higher spatial dimensions to observe its effect, or 
non-local interactions; indeed, the relevance of exchange statistics 
is related to the connectivity of the lattice. The second comment is the
observation that similar to the non-local rotation of statistics
$\theta$ one could define local rotations between different flavors 
\cite{our4}. 
  
\bibitem{Note2}
In general (since there is a continuous $SU(2)$ symmetry, apart from
the $U(1)$, that is broken) when ${\cal N}_\uparrow$ and  ${\cal
N}_\downarrow$ are both order $N_s$, there will be two eigenvectors
with eigenvalues of order $N_s$ and, therefore, the condensate is a
mixture. 

\bibitem{our2}
C.D. Batista and G. Ortiz, \Journal{\PRL}{85}{4755}{2000}.

\end{thebibliography}
\end{document}